\documentclass[conference, letterpaper]{IEEEtran}
\IEEEoverridecommandlockouts
\usepackage{cuted}
\usepackage{cite}
\usepackage{bm}
\usepackage{amsmath,amssymb,amsfonts}
\usepackage{algorithm}
\usepackage{lettrine}
\usepackage{graphicx}
\usepackage{multirow}
\usepackage{algpseudocode}
\usepackage{diagbox}
\usepackage{graphicx}
\usepackage{subcaption}
\usepackage{color,soul}
\usepackage[letterpaper, left=0.625in, right=0.625in, bottom=1.02in, top=0.70in]{geometry}
\usepackage{slashbox}
\usepackage{textcomp}
\usepackage{xcolor}

\newcommand{\be}{\begin{equation}}
\newcommand{\ee}{\end{equation}}

\def\BibTeX{{\rm B\kern-.05em{\sc i\kern-.025em b}\kern-.08em
    T\kern-.1667em\lower.7ex\hbox{E}\kern-.125emX}}

\begin{document}

%\title{Model-based Deep Learning for Rate Split Multiple Access in MISO Communications System}
\title{Model-based Deep Learning for Wireless Resource Allocation in RSMA Communications Systems}

\author{\IEEEauthorblockN{Hanwen Zhang\IEEEauthorrefmark{1}, 
Mingzhe Chen\IEEEauthorrefmark{2},
Alireza Vahid\IEEEauthorrefmark{3},
Feng Ye\IEEEauthorrefmark{4},
Haijian Sun\IEEEauthorrefmark{1}}
\IEEEauthorblockA{
\IEEEauthorrefmark{1}School of Electrical and Computer Engineering, University of Georgia, Athens, GA, USA \\
\IEEEauthorrefmark{2}Department of Electrical and Computer Engineering,
University of Miami, Coral Gables, FL, USA\\
\IEEEauthorrefmark{3}Electrical and Microelectronic Engineering, Rochester Institute of Technology, Rochester, NY, USA\\
\IEEEauthorrefmark{4}Department of Electrical and Computer Engineering, University of Wisconsin-Madison, Wisconsin, WI, USA
\\
Emails: hanwen.zhang@uga.edu, mingzhe.chen@miami.edu, arveme@rit.edu, feng.ye@wisc.edu, hsun@uga.edu}}
\maketitle

\begin{abstract}
Rate-splitting multiple access (RSMA) has been proven as an effective communication scheme for 5G and beyond. However, current approaches to RSMA resource management require complicated iterative algorithms, which cannot meet the stringent latency requirement by users with limited resources. Recently, data-driven methods are explored to alleviate this issue. However, they suffer from poor generalizability and scarce training data to achieve satisfactory performance. In this paper, we propose a fractional programming (FP) based deep unfolding (DU) approach to address resource allocation problem for a weighted sum rate optimization in RSMA. By carefully designing the penalty function, we couple the variable update with projected gradient descent algorithm (PGD). Following the structure of PGD, we embed a few learnable parameters in each layer of the DU network.
Through extensive simulation, we have shown that the proposed model-based neural networks can yield similar results compared to the traditional optimization algorithm for RSMA resource management but with much lower computational complexity, less training data, and higher resilience to out-of-distribution (OOD) data. 
\end{abstract}

%\begin{IEEEkeywords}
%RSMA, model-based deep learning, deep unfolding, projection gradient descent, low complexity, fractional programming
%\end{IEEEkeywords}

\section{Introduction}
% RSMA intoduction
Rate-splitting multiple access (RSMA) is a cutting-edge multiple access technique in future wireless communication systems \cite{clerckx2023primer,clerckx2016rate,yang2021optimization,yang2020sum,xiao2023joint,dizdar2023rsma}. RSMA divides the data stream into a common part that conveys information shared by multiple users, and a private part that fulfills individual user requirements. For example, in vehicular communication, the common message can be critical safety information and private messages may include each vehicle's control and personnel data \cite{ma2024exploring}. RSMA efficiently utilizes spectrum resources by broadcasting the common information, thereby freeing up bandwidth for private data and improving individual quality-of-service (QoS). In \cite{yang2021optimization}, it derived a closed-form solution for the optimal private beamformer and showed that RSMA has advantages in resource allocation tasks with a massive number of users, minimum rate demand of users, and low transmit power scenarios. It also provided a basic scheme for resource allocation in single-input single-output (SISO) RSMA systems. Other works such as \cite{clerckx2023primer,clerckx2016rate,clerckx2024multiple} demonstrated RSMA robustness, especially in scenarios with high latency sensitivity. To reduce system latency in wireless communications, a lower computational workload is also required~\cite{xiao2023joint,dizdar2023rsma}. Specifically, \cite{xiao2023joint} proposed a low complexity algorithm by reducing the redundant constraints and pre-processing deployment region modeling. \cite{dizdar2023rsma} designed zero forcing (ZF) and maximum ratio transmission (MRT) based methods to find closed-from expression solution for the RSMA max-min fairness problem and reduce computational complexity. However, these existing low complexity algorithms can barely meet the stringent latency requirements in high mobility scenarios.

% learning-> model-based
Deep learning has emerged as a pivotal approach to enhance the performance of communication systems \cite{zhang2022deep,gao2022online}. Leveraging large datasets, deep learning facilitates the development of neural network models that are capable of learning from data implicitly \cite{zhang2023map2schedule}. However, the mappings learned during the training process often yield unpredictable outcomes when exposed to out-of-distribution (OOD) scenarios, which is caused by the lack of explicit physical principles constraints \cite{shlezinger2023model}. To address this problem, high volume and diverse data are required to train a model that is able to make neural networks robust in OOD scenarios. Recently, model-based deep learning is considered a promising alternative to problems where the paradigm incorporates domain knowledge to design an interpretable neural network architecture. It enables the training of neural networks with less data while achieving robust performance in OOD settings \cite{shlezinger2023model}.  In the context of wireless resource allocation, several studies have proposed learning-based optimization strategies. For instance, \cite{schynol2023coordinated} presented a graph convolution network weighted minimum mean square error (GCN-WMMSE) deep unfolding (DU) algorithm that combines GCNs with WMMSE method to achieve robust neural networks based resource allocation. Meanwhile, \cite{zhang2023joint} adopted Uzawa's method \cite{gallier2019fundamentals} to integrate constraint-related variables with the objective function for DU network design. However, above works usually consider simple constraints in their resource allocation problems, and fail to include important communication-related metrics.  

Until recently, the work in \cite{xia2023deep} extended resource allocation to general scenarios, using a fractional programming (FP) based DU framework to tackle the weighted sum rate (WSR) problem in RIS scenario under data rate constraints. But the penalty function used in this work is approximated by a neural network, which limited their generalizability. Motivated by this, 
in our work, we focus on model-based deep learning to solve a resource allocation problem in multiple-input single-output (MISO) RSMA with common rate constraints.  We applied the FP framework and derived the projected gradient descent (PGD) updating for the entire problem. Moreover, the specially designed learnable parameters can specifically enhance robustness in OOD test with much lower complexity. We summarize the contributions of this paper as follows:
\begin{itemize}
\item We address a multi-user MISO RSMA resource allocation utilizing FP-based DU. This FP-based DU is able to learn the converge trajectory from a small mount of data. Thus, the resource allocation task computation complexity is reduced while performance still remains.

\item The FP framework is extended by applying the PGD algorithm and unfolding structure. To address the updating problem of variables that only exist in the constraint, we introduce a specially designed penalty factor to update the common stream beamformer in PGD. Subsequently, to boost the algorithm performance, the DU-based interpreted learning framework is specially designed with only few learnable parameters. 

\item Finally, the experiments provide a comprehensive analysis on the convergence to show proposed algorithm has similar performance as traditional optimization. Then, we provide the OOD test result to validate that the structure we designed is robust in different environments. The last experiment shows the proposed model has much lower computational complexity than traditional algorithms. The provided experiments sufficiently show model-based solution has the potential for wireless resource allocation optimization.
\end{itemize}

The rest of the paper is organized as follows. In Section \ref{section_II}, the multi-users MISO RSMA system model is presented. The corresponding traditional algorithm and DU are detailed in Section \ref{section_III}. In Section \ref{section_IV}, experiments evaluate DU's performance, generalization ability and computation complexity. Finally, Section \ref{section_V} concludes the paper.

Notation: We use bold capital symbol as matrix, bold lowercase symbol as vectors, the lowercase symbol as scalar, $||.||$ as $l_2$ norm, $\text{Re}\{.\}$ as real part of function, $[.]^H$ as Hermite transpose, $[.]^T$ as transpose, $\mathcal{U}\{.\}$ as uniform distribution, $\mathcal{N}\{.\}$ as Gaussian distribution, $\Bar{z}$ as conjugate of $z$.
\section{System Model}\label{section_II}
\subsection{System Model}

We consider a downlink multi-user communication system, which consists of a single base station (BS) with $M$ antennas and a total of $U$ single-antenna users. In an RSMA system, data is split into common and private streams. The common stream embeds messages for all users while the private stream provides individual data for each user. Denote $\mathbf{v}_0,\mathbf{v}_k \in \mathbb{C}^{M\times 1} $ the beamforming vectors for common and private data streams, respectively. The BS transmit signal for common stream and user $k$ private stream are given by:
\begin{flalign}                     \mathbf{x}_0&=\mathbf{v}_0s[n], \\
        \mathbf{x}_k&=\mathbf{v}_ks_k[n], k\in\{1,2,..,U\},
\end{flalign}
where $s[n]$ and $s_k[n]$ are the common and dedicated data streams, respectively. $E(s[n])^2 = E(s_k[n])^2 =1$, i.e., both data streams are normalized. The received signal of user $k$ is:
\begin{flalign}
    {y}_k=\mathbf{h}_k^H\mathbf{x}_0 +\mathbf{h}_k^H\mathbf{x}_k+\sum_{j\neq k}^U \mathbf{h}_k^H\mathbf{x}_j + {n}_k,
\end{flalign}
where $\mathbf{h}_k \in \mathbb{C}^{M\times 1}$ is the channel between BS and user $k$, which follows circularly symmetric complex Gaussian distribution with power $h_k$, $\mathbf{h}_k \sim \mathcal{CN}(\mathbf{0},h_k^2\mathbf{I})$. ${n}_k\sim \mathcal{N} (0,\sigma_k^2) $ is AWGN noise. The achievable common stream data rate of user $k$ is given below in (\ref{comm_stream}), while the private stream is given as (\ref{private_stream}):
\begin{flalign}
    &c_k=\log_2(1+\frac{\mathbf{h}_k^H \mathbf{v}_0\mathbf{v}_0^H\mathbf{h}_k}{\sigma_k^2+\sum_{j =1}^U \mathbf{h}_k^H \mathbf{v}_j\mathbf{v}_j^H\mathbf{h}_k}),\label{comm_stream}\\
    &R_k^p=\log_2(1+\frac{\mathbf{h}_k^H \mathbf{v}_k\mathbf{v}_k^H\mathbf{h}_k}{\sigma_k^2+\sum_{j \neq k}^U \mathbf{h}_k^H \mathbf{v}_j\mathbf{v}_j^H\mathbf{h}_k}).\label{private_stream}
\end{flalign}

To ensure that every user is able to decode the common data stream, the lowest common data stream channel capacity must be larger than the sum of user's common stream decoding rate. Therefore, it has the rate limitation given by \cite{clerckx2023primer,clerckx2016rate}
\begin{flalign}
    \sum_{k=1}^U R_k^c\leq\min(c_k),
\end{flalign}
where $\min(c_k)$ is the lowest common stream rate among all users, $R_k^c$ is the common data rate allocated to $k$-th user. Without loss of generality, we assume user 1 has the expected lowest channel gain, the sum of common stream rate upper bound is given by \cite{yang2021optimization}:
\begin{flalign}
    \min(c_k) = \log_2(1+\frac{\mathbf{h}_1^H \mathbf{v}_0\mathbf{v}_0^H\mathbf{h}_1}{\sigma_k^2+\sum_{k=1}^U \mathbf{h}_1^H \mathbf{v}_k\mathbf{v}_k^H\mathbf{h}_1}).
\end{flalign}
\subsection{Problem Formulation}
To evaluate the RSMA system performance, we apply WSR as the performance metric. Accordingly, the downlink RSMA WSR problem is formulated as: 
\begin{subequations}
    \begin{flalign}
        \mathbf{P}_1 :&\max_{\mathbf{v}_k,R_k^c}   \text{WSR} =  \sum_{k=1}^U f_k(R_k^c+R_k^p), \label{WSR_org}\tag{8}\\
        \text{s.t.}\
        &\sum_{k=0}^U \mathbf{v}_k^H \mathbf{v}_k+\text{P}_c \leq \text{P}_\text{max},\label{totalpowerconstraints}\\
        &\sum_{k=1}^U R_k^c\leq \min(c_k), \label{maxRcsum}\\
        &\text{Tr}(\mathbf{v}_k\mathbf{v}_k^H)\geq \text{P}_0, \label{minBFpower}\\
        &R_k^c\geq 0,\label{Rkc_exist}
    \end{flalign}\label{P1}%
\end{subequations}
where (\ref{WSR_org}) is the original problem in which the objective is to maximum the system WSR. $f_k$ is the weight for user $k$; (\ref{totalpowerconstraints}) sets the total power consumption to be lower than BS power $\text{P}_\text{max}$, and $\text{P}_c$ is fixed circuit power consumption; in (\ref{maxRcsum}), it guarantees common stream can be decoded by every user; constraint (\ref{minBFpower}) guarantees the users' private data transmission quality by setting beamformers power lower bound; (\ref{Rkc_exist}) ensures the existence of common stream.

\section{Proposed Solution}\label{section_III}
In this section, we solve the original problem $\textbf{P}_1$ by the iterative FP algorithm \cite{shen2018fractional}. Then, we design a problem with a penalty factor to match PGD updating requirement. By adding learnable parameters and applying PGD algorithm, we propose a FP based DU method to address the problem with low complexity and high robustness. 
\subsection{Fractional Programming}
We apply the standard semi-definite relaxation (SDR) and let $\mathbf{V}_k=\mathbf{v}_k\mathbf{v}_k^H$. Besides, we define auxiliary variables: 
\begin{flalign}
  &z_k^*=(\sigma_k^2+\sum_{j\neq k}^U \mathbf{h}_k^H \mathbf{V}_j\mathbf{h}_k)^{-1}\mathbf{h}_k^H\mathbf{v}_k. \label{z_k_update}\\
  &z_0^{*}=(\sigma_k^2+\sum_{ k=1}^U \mathbf{h}_1^H \mathbf{{V}}_k\mathbf{h}_1)^{-1}\mathbf{h}_1^H\mathbf{{v}}_0^{i}\label{z0_update},
\end{flalign}
Given initial feasible values for $\mathbf{v}_k$, $z_k^*$ and $z_0^*$ would be constant.  And we set $\Phi_{0}=1+2\text{Re}\{\sqrt{\bar{z}_0\mathbf{h}_1^H \mathbf{{V}}_0\mathbf{h}_1 z_0}\}-\bar{z}_0(\sigma_k^2+\sum_{k=1}^U\mathbf{h}_1^H \mathbf{{V}}_k \mathbf{h}_1 )z_0$ and $\Phi_{k}=1+2\text{Re}\{\sqrt{\bar{z}_k\mathbf{h}_k^H \mathbf{{V}}_0\mathbf{h}_k z_k}\}-\bar{z}_k(\sigma_k^2+\sum_{j\neq k}^U\mathbf{h}_k^H \mathbf{{V}}_j\mathbf{h}_k )z_k$.
Then, from FP principle \cite{shen2018fractional}, $\textbf{P}_1$ can be equivalently transformed as
\begin{subequations}
    \begin{flalign}
        \mathbf{P}_2 :&\max_{\mathbf{V}_k,R_k^c}  \sum_{k=1}^U f_k \big(R_k^c+\log_2 (\Phi_k)\big), \tag{11}\\
        \text{s.t.}\
        &\sum_{k=1}^U R_k^c\leq \log_2(\Phi_0), \label{maxRcsum_FP}\\
        &(\ref{totalpowerconstraints}),(\ref{minBFpower}),(\ref{Rkc_exist}),\notag
    \end{flalign}
    \label{P2}%
\end{subequations}
It can be readily shown that given $z_k^*$, $\mathbf{P}_2$ is convex with respect to $\mathbf{v}_k$ and $R_k^c$. 
Therefore, it can be solved by well-known toolbox, such as \text{CVX} \cite{grant2014cvx}. Then $z_k^*$ is updated by $\mathbf{P}_2$'s solution $\mathbf{v}_k$. This iterative algorithm is shown in \text{Algorithm \ref{algorithm_FP_1}}. 
%However, this algorithm has significant computational complexity, as shown in later sections. Thus, we are motivated to identify a learning-based approach for a more efficient solver. 

\begin{algorithm}
    \caption{FP Beamforming for WSR optimization}\label{algorithm_FP_1}
    \begin{algorithmic}
        \Require  $\mathbf{h}_k$, $f_k$,  $P_0$, $P_c$, $P_\text{max}$, initial value of $z_0$, $z_k$, $\mathbf{v}_0$, $\mathbf{v}_k$ and $R_k^c$. Set counter $j=1 $ and convergence precision $\phi_p$. 
        \While{$|\text{WSR}_{j+1}-\text{WSR}_{j}| > \phi_p$}
            \State\textbf{Step 1} Update $\mathbf{V}_k$ and $\text{WSR}_{j}$ from $\mathbf{P}_2$ with ${z}_k^*$, ${z}_k^*$,
            \State\textbf{Step 2} Apply eigen decomposition on $\mathbf{V}_k$ to obtain $\mathbf{v}_k$,
            \State\textbf{Step 3} Update each $z_k^*$, $z_0^*$ by (\ref{z_k_update}) and (\ref{z0_update}), $j=j+1$.
        \EndWhile%
    \end{algorithmic}%
\end{algorithm}%

\subsection{Projection Gradient Descent with A Penalty Function}
To gain more insights, we will derive semi-closed-form solutions in this subsection. Specifically, the PGD algorithm is introduced, which splits $\mathbf{P}_2$ into two parts: the gradient descent on the objective function (unconstrained problem) and a projection to ensure all constraints are met. Since in PGD algorithm, the derivative is directly based on $\mathbf{v}_k$ and $\mathbf{v}_0$ instead of SDR variable $\mathbf{V}_k$ and $\mathbf{V}_0$.
However, simply dropping all constraints in $\mathbf{P}_2$ will lead to no gradient updates for $\mathbf{v}_0$. Therefore, we first reformulate $\textbf{P}_2$ by adding a penalty function of constraint (\ref{maxRcsum}) to the objective function, which couples gradient with $\mathbf{v}_0$. The unconstrained problem then becomes:
\begin{flalign}
   \mathbf{P}_3:
    \max_{\mathbf{\Tilde{v}}_k,\Tilde{R}_k^c}   & \mathcal{L}=\sum_{k=1}^U f_k\Big(\Tilde{R}_k^c+\log_2\big(\Phi_k) \Big)\notag\\
    &-\lambda(\sum_{k=1}^U \Tilde{R}_k^c-\log_2(\Phi_0)). 
    \label{P3}%
\end{flalign}%
Here, $\lambda > 0$ is the penalty factor. When PGD updates $\mathbf{v}_k, \forall k$ in RSMA system, $\mathbf{v}_0$ can be updated in each iteration with the penalty function given in $\mathbf{P}_3$. Therefore, based on the unconstrained convex optimization in $\mathbf{P}_3$, the beamformers and common data rate $\mathbf{\Tilde{v}}_k^{i+1},\mathbf{\Tilde{v}}_0^{i+1}, \Tilde{R}_k^{c,{i+1}}$  can be  updated by $\mathbf{{v}}_k^i,\mathbf{{v}}_0^i, {R}_k^{c,i}$ following their gradient descent direction. The iterative steps are given in (\ref{update_pgd}), respectively. 
\begin{flalign}
    &\Tilde{R}_k^{c,i+1}=R_k^{c,i}+\alpha_{1,k}\nabla{R}_k^{c,i}\notag,\\
    &\mathbf{\Tilde{v}}_0^{i+1}=\mathbf{v}_0^{i}+\alpha_2\nabla\mathbf{{v}}_0^{i},\notag\\
    &\mathbf{\Tilde{v}}_k^{i+1}=\mathbf{v}_k^{i}+\alpha_{3,k}\nabla\mathbf{{v}}_k^{i},\label{update_pgd}
\end{flalign}
where $i$ is $i$-th iteration step, $\alpha_{1,k}$, $\alpha_2$ and $\alpha_{3,k}$ are fixed step size. And $\nabla{R}_k^{c,i},\nabla\mathbf{{v}}_0^{i}, \nabla\mathbf{{v}}_k^{i}$ are gradients (derivatives) of $\mathcal{L}$ with respect to $R_k^c$, $\mathbf{v}_0$, and $\mathbf{v}_k$ are given as
\begin{flalign}
 &\nabla{{R}}_k^{c,i}=\frac{\partial \mathcal{L}}{\partial {R}_k^{c,i}}=f_k-\lambda,\forall k \neq 0,\\
    &\nabla\mathbf{{v}}_0^{i}=\frac{\partial \mathcal{L}}{\partial \mathbf{v}_0^i}=\frac{2\lambda\bar{z}_0^{i}\mathbf{h}_1}{\Phi_0^i\ln2},\\
    &\nabla\mathbf{{v}}_k^{i}=\frac{\partial \mathcal{L}}{\partial \mathbf{v}_k^i}=(\frac{\bm{\zeta}_k^i}{\Phi_k^i}+\sum_{j\neq k}^U\frac{\bm{\beta}_{j,k}^i}{\Phi_j^i}+ \mathbf{o}_{k}^i),\forall k \neq 0,\label{Vk_update_org}
\end{flalign}
where $\bm{\zeta}_k^i$, $\bm{\beta}_{j,k}^i$ and $\mathbf{o}_{k}^{i}$ are given as
\begin{flalign}
    &\bm{\zeta}_k^i=\frac{\partial (2f_k\text{Re}\{z_k^{i} \mathbf{h}_k^H \mathbf{{v}}_k^i\})}{\partial \mathbf{v}_k^i\ln2}={{2f_k \bar{z}_k^{i} \mathbf{h}_k }}/\ln2,\\
    &\bm{\beta}_{j,k}^i=-\frac{\partial f_j\bar{z}_j^{i}(\sigma_k^2+\sum_{l\neq j}^U\mathbf{h}_j^H \mathbf{{v}}_l^{i} \mathbf{{v}}_l^{i,H} \mathbf{h}_j )z_j^{i}}{\partial \mathbf{v}_k^i\ln2}\notag\\
    &=-2 z_j^{i} \bar{z}_j^{i}f_j\mathbf{h}_j  \mathbf{h}_j^H\mathbf{{v}}_k^{i}/\ln2,\forall k \neq 0,\\
    &\mathbf{o}_{k}^{i}=\frac{\partial\lambda\left(\sum_{k=1}^U -{R}_k^{i,c}+\log_2(\bm{\Phi}_0^i)\right)}{\partial \mathbf{{v}}_k^{i}}= \frac{-2 \lambda z_0^i \bar{z}_0^i\mathbf{h}_1\mathbf{h}_1^H \mathbf{{v}}_k^{i}}{\Phi_0^i\ln2}.\label{Pn_update_org}
\end{flalign}

To guarantee the updated variables  in the feasible region of the original problem, we employ a projection method to meet all constraints in $\mathbf{P}_2$. Hence, the projection algorithm for $\mathbf{\Tilde{v}}_k^{i}, \forall k \in \{0,1,...,U\}$ is given as:
\begin{flalign}
    &\mathbf{v}_k^{i+1} = \frac{\mathbf{\Tilde{v}}_k^{i+1}}{||\mathbf{\Tilde{v}}_k^{i+1}||}\sqrt{a_k^*},k=\{0,1,...,U\},\notag\\
     &a_k^*=\frac{\Tilde{a}_k}{\sum_{k=0}^U \Tilde{a}_k}(\text{P}_\text{max}-(U+1)\text{P}_0)+\text{P}_0,\notag\\
    &\Tilde{a}_k=\max(||\mathbf{\Tilde{v}}_k^{i+1}||_2^2,\text{P}_0)-\text{P}_0.\label{projectionv_k}
\end{flalign}

The projection approach above bounds the beamformer power within feasible region without changing beamforming direction. Moreover, the corresponding projection for $\Tilde{R}_k^{c,i+1}$ is given as
\begin{flalign}
    &R_k^{c,i+1}=\frac{\max(\Tilde{R}_k^{c,i},0)}{\sum_{k=1}^U\Tilde{R}_k^{c,i}}\min(c_k^{i+1})\notag\\
    &=\frac{\max(\Tilde{R}_k^{c,i},0)}{\sum_{k=1}^U\Tilde{R}_k^{c,i}}\log_2(1+\frac{\mathbf{h}_k^H \mathbf{v}_k^{i+1}\mathbf{v}_k^{i+1,H}\mathbf{h}_k}{\sigma_k^2+\sum_{j\neq k}^U \mathbf{h}_k^H \mathbf{v}_j^{i+1}\mathbf{v}_j^{i+1,H}\mathbf{h}_k}).
\end{flalign}

When the number of iteration approaches to infinity, $R_k^{c*}$, $k\in\{k|f_k= \max(f_1,f_2,...,f_U)\}$, has asymptotic property to achieve upper bound, $\min (R_k)$, while other $R_k^c=0$. Therefore, projection step for $R_k^c$ is simplified as: 
\begin{equation}
R_k^{i+1,c} = 
\begin{cases} 
\min(c_k^{i+1}), & \text{if } f_k = \max(f_1,f_2,...,f_U),\\
0, & \text{otherwise}.
\end{cases}\label{R_kc_map}
\end{equation}
\subsection{PGD Based Deep Unfolding}
The gradient updating step given in above PGD, however, has poor convergence performance, particularly when the step size is small. Moreover, the penalty function given in $\mathbf{P}_3$ even punishes the objective function when the constraints are in feasible region. Therefore, in this subsection, we design a DU structure \cite{shlezinger2023model}, by adding learnable parameters $\bm{\theta}$ into the PGD gradient updating part. These learnable parameters are trained to dynamically adjust weights for beamformers at each iteration, thereby can converge within few steps, which is much faster than PGD. 

Essentially, the  DU structure is designed to follow each iteration in PGD. The difference is that, DU introduces some learnable parameters which assign the weights for the terms in polynomials of $\Pi_1$ and $\Pi_2$ to update $\mathbf{v}_0^i$ and $\mathbf{v}_k^i$. Specifically,  in the $n$-th layer,  $\mathbf{v}_0$ and $\mathbf{v}_k$ in (\ref{update_pgd}) are reformulated as:
\begin{flalign}
&\mathbf{\Tilde{v}}_0^{n+1}=\mathbf{v}_0^{n}+\Pi_1(\mathbf{{v}}_0^{n}),\label{v0_update_simple}\\
    &\mathbf{\Tilde{v}}_k^{n+1}=\mathbf{v}_k^{n}+\Pi_2(\mathbf{{v}}_k^{n}).\label{vk_update_simple}
\end{flalign}

Here, $\Pi_1(\mathbf{{v}}_0^{n})$ and $\Pi_2(\mathbf{{v}}_k^{n})$ are the $n$-th layer for  $\mathbf{v}_0$ and $\mathbf{v}_k$, which are given as,

\begin{flalign}
    &\Pi_1(\mathbf{{v}}_0^{n})=\ln2 (\bm{\phi}\mathbf{w}_0^n)\frac{\nabla\mathbf{{v}}_0^{n}}{\lambda},\label{V0_update_pi}\\
    &\Pi_2(\mathbf{{v}}_k^{n})=(\bm{\phi}\mathbf{w}_k^n)
    \begin{bmatrix}\frac{\bm{\zeta}_k^n}{{\Phi_k^n}},\frac{\bm{\beta}_{1,k}^n}{{{\Phi_1^n}}},\frac{\bm{\beta}_{2,k}^n}{{{\Phi_2^n}}},...\frac{\bm{\beta}_{j,k}^n}{{{\Phi_j^n}}},..,\frac{\bm{\beta}_{U,k}^n}{{{\Phi_U^n}}},\frac{\mathbf{o}_{k}^n}{\lambda}
    \end{bmatrix}\notag\\
    &\begin{bmatrix} \eta_k^n,\eta_j^{n,1},\eta_j^{n,2},...,\eta_j^{n,j},...,\eta_j^{n,U},\eta_k^{n,p}
    \end{bmatrix}^T\ln2, \forall k \neq 0,\forall j \neq k,\label{Vk_update_pi}\\
    &\bm{\phi}=[f_1,f_2,...,f_U,\text{P}_0,\text{P}_\text{max}],%
\end{flalign}
where $\bm{\phi}$ is the environment pattern vector, which is spliced by weights $f_k$ and power consumption $\text{P}_\text{max},\text{P}_0$. And the learnable parameters for each layer are $\bm{\theta}^{n}=\{\mathbf{w}_0^{n,T},\mathbf{w}_k^{n,T}, \eta_k^{n}, \eta_j^{n,k} ,\eta_k^{n,p}\}$,
where $\mathbf{w}_0^{i}, \mathbf{w}_k^{n}\in \mathbb{R}^{(U+2)\times 1}$,  $\eta_k^{n,p}, \eta_j^{n,k}, \eta_k^{n} \in \mathbb{R}$. 
% To ensure that all variables are updated following the gradient descent direction in $\Pi_1$ and $\Pi_2$, we regulate all parameters to be greater than $\epsilon$, as in $\bm{\theta}^{n}\succ \epsilon$, where $\epsilon$ is a small positive scalar.{globecom2024_00.png} 

\begin{figure*}[!h]
    \centering
    \includegraphics[width=5.5in]{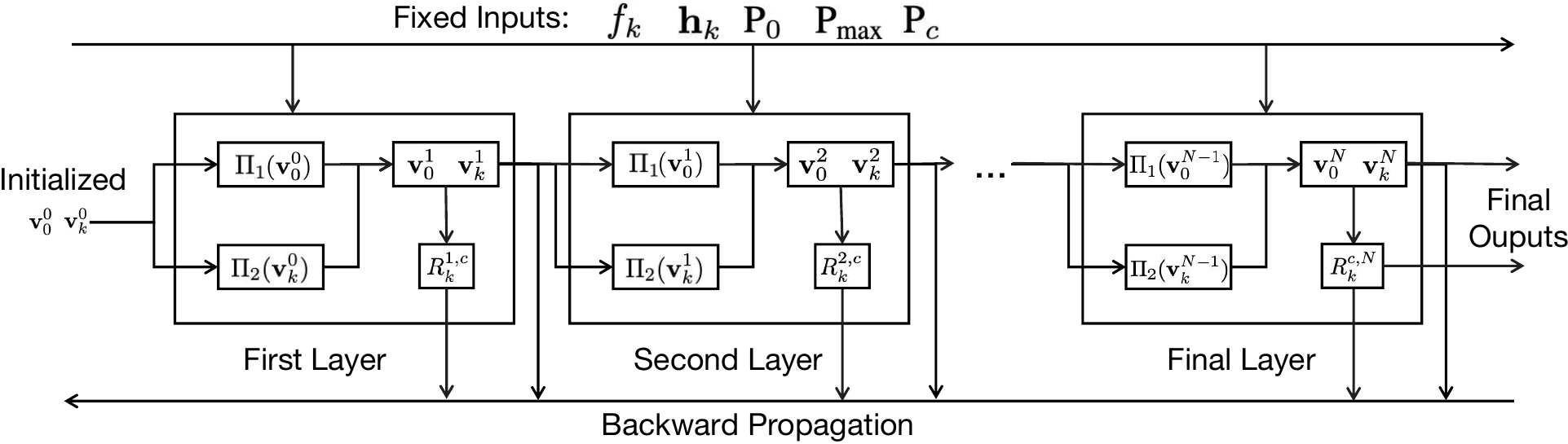}
	\centering
	\caption{Proposed Deep Unfolding Networks Structure Overview}
	\label{fig:RSMA_DU_model}
\end{figure*}

The designing of learnable parameters follows some schemes proposed in \cite{samuel2019learning,nguyen2020deep}, which add parameters on each term in gradient polynomial to change the weights of updating in different directions. Besides, in this particular problem, the penalty function given above has penalty factor $\lambda$. If we consider the penalty factor as the Lagrangian multiplier, the training process can be considered as the regression problem for learnable parameter $\eta_k^p$ to regress the Lagrangian multiplier value in each iteration step. The learnable parameter, $\mathbf{W}_k$, is applied to construct a linear transformation with $\bm{\phi}$. This linear transformation provides a robust factor to make PGD be adaptive to $f_k$, $\text{P}_0$ and $\text{P}_\text{max}$. Therefore, the beamformers gradient descent is able to be accelerated specifically with $\phi$.  DU algorithm is given as \text{Algorithm \ref{algorithm_DU}}. A graphic illustration of DU layers is shown in Fig. \ref{fig:RSMA_DU_model}.

\begin{algorithm}
    \caption{Deep Unfolding for FP-based WSR Optimization}\label{algorithm_DU}
    \begin{algorithmic}
        \Require  Initialize hyper-parameters and data,\\
        $\text{counter}=0$,
        \While{$\text{epoch} - \text{counter}>0$}\\
        $\text{counter}=\text{counter}+1$,\\        
        \textbf{Forward Propagation Part:}
        
            \State Initial beamformers randomly, then for each layer:
            \State\textbf{Step 1} Update $z_k$, by (\ref{z_k_update}); Update $z_0$, by (\ref{z0_update}),
            \State\textbf{Step 2} Update $\mathbf{\Tilde{v}}_0$, by (\ref{v0_update_simple}); Update $ \mathbf{\Tilde{v}}_k$, by (\ref{vk_update_simple}),
            \State\textbf{Step 3} Update $ \mathbf{{v}}_k$ and $ \mathbf{{v}}_0$ by projection (\ref{projectionv_k}), $ R_k^c$ by (\ref{R_kc_map}),
            \State\textbf{Step 4} Output results $\{\mathbf{v}_0,\mathbf{v}_k,R_k^c\}$,\\
        \textbf{Forward Propagation ends} \\
        \textbf{Backward Propagation Part:}
            \State\textbf{Step 5} Calculate the $\text{loss}$ given by (\ref{loss_func}),
            \State\textbf{Step 7} Backward Propagation, update $\bm{\theta}$ and optimizer.
            % \State\textbf{Step 8} Regulate parameters to be larger than $\epsilon$ by \\$\bm{\theta}^n=\max(\bm{\theta}^n,\epsilon$),\\
        \textbf{Backward Propagation ends}  
        \EndWhile
    \end{algorithmic}
\end{algorithm}

Lastly, for this learning problem, the loss function is: 
\begin{flalign}
    &\text{Loss}=\frac{1}{QN}\sum_{q=1}^Q\sum_{n=1}^N\log_2(n+1)(\hat{\text{WSR}}_{q,n}-\text{WSR}_{q,n}^*),\label{loss_func}
\end{flalign}
\begin{flalign}
    &{\hat{\text{WSR}}_{q,n}}=\notag\\
    &\sum_{k=1}^U f_k^q({R}_{k,q}^{c,n}+\log_2(1+\frac{\mathbf{h}_{k,q}^H \mathbf{{v}}_{k,q}^n\mathbf{{v}}_{k,q}^{n,H}\mathbf{h}_{k,q}}{\sigma_{k,q}^2+\sum_{j\neq k}^U \mathbf{h}_{k,q}^H \mathbf{{v}}_{j,q}^n\mathbf{{v}}_{j,q}^{n,H}\mathbf{h}_{k,q}})),%
\end{flalign}
where $q$ is the $q$-th sample in $Q$ samples batch. $\text{WSR}^*$ is the ground truth WSR provided in dataset, generated by \text{Algorithm \ref{algorithm_FP_1}}. The log loss function is similar to the shape of the iterative step. As a result, loss function couples the relationships between last layer outputs and each layer outputs like iterative process in traditional algorithm. Besides, the loss function sets a connection between each layer and loss function output to prevent gradient vanishing.

\section{Simulation Results}\label{section_IV}
In this section, we evaluate the performance of proposed DU networks via numerical simulations. Firstly, we introduce evaluation settings on dataset and training hyper-parameters. For experiments, the convergence evaluation is given to show which set of hyper-parameters gets the best results. Then, we test the robustness on OOD data set with different distributions of channel SNR and $\text{P}_\text{max}$. Finally, the complexity comparison is given to show the proposed algorithm has much faster computation speed than traditional algorithm. 

\subsection{Evaluation Settings and Training Hyper-Parameters}

% parameters setting
The evaluation considers a scenario with $U=3$ users in downlink RSMA MISO system. The BS has $M=12$ antennas with a maximum transmission power $\text{P}_\text{max} = 33  \text{ dBm}$. The circuit power consumption is assumed $\text{P}_\text{c}=30 \text{ dBm}$. Channel state information is represented as CSCG, where $\mathbf{h}_k\sim \mathcal{CN}(0,10\mathbf{I})$, ${n}_k\sim \mathcal{N}(0,1)$. The convergence precision for FP IPM approach is $\phi_p = 10^{-2}$. The weights $f_k$, minimum power requirement of $\mathbf{v}_k$ and AWGN power are stochastically generated following a uniform distribution, i.e, ${f}_k\sim \mathcal{U}(0,1)$ while $\sum_{k=1}^U{{f}_k}=1$,  $\text{P}_0\sim \mathcal{U}(0,0.125)$. The dataset has $3,000$ samples generated by \text{Algorithm 1} using CVX tool box. Unless otherwise stated, we choose $N=8$, $lr=0.003$ as optimizer learning rate, $bz=1000$ as batch size, and $ep=480$ as epoch for the hyper-parameters setting.

% \subsection{Important Parameters Comparison and Effectiveness Test}
% \begin{table*}[ht]
% \centering
% \caption{Important Parameters Comparison in Test Set Performance}
% \begin{tabular}{|c|c|c|c|c|c|c|c|c|c|c|c|} 
% \hline
% \multirow{2}{*}{Parameters}& \multicolumn{5}{c|}{Learning Rate Part Test Set Performance} &  \multicolumn{5}{c|}{Batch Size Part Test Set Performance}   \\ 
% \cline{2-11} 
%  & $lr=0.015$ & $lr= 0.01$ & $lr=0.005 $& $lr=0.003 $& $lr=0.001 $& $bz=200 $& $bz=400 $& $bz=700 $& $bz=1000 $&$bz=2000 $\\
% \hline
% ASR (\%) & $96.77$ & $\mathbf{96.99}$ & $96.78$ & $96.71$ & $95.18$ & $\mathbf{97.08}$ & $96.88$&$96.66$ & $96.71$& $95.36$\\ 
% \hline
% \multirow{2}{*}{Parameters}& \multicolumn{10}{c|}{Layer Number Part Test Set Performance} \\
% \cline{2-11}
% & \multicolumn{2}{c|} {$N=4$} & \multicolumn{2}{c|} {$N= 6$}& \multicolumn{2}{c|} {$N= 8$} & \multicolumn{2}{c|} {$N=10$} & \multicolumn{2}{c|} {$N=12$} \\
% \hline
% ASR (\%) & \multicolumn{2}{c|}{$85.75$} & \multicolumn{2}{c|}{$92.50$} & \multicolumn{2}{c|}{$94.70$} & \multicolumn{2}{c|}{$96.39$} & \multicolumn{2}{c|}{$\mathbf{96.71}$}\\
% \hline
% \end{tabular}

% \label{table:important_parameters_comparison}
% \end{table*}

% In this subsection,  Specifically, learning rate, batch size, and layer number are compared to find the best hyper-parameters and prove the effectiveness of the proposed method.  %By control variate method, we test the impact of different hyper-parameter values.{test1_00.png}

\begin{figure}[!h]
    \centering
     \includegraphics[width=2.9in]{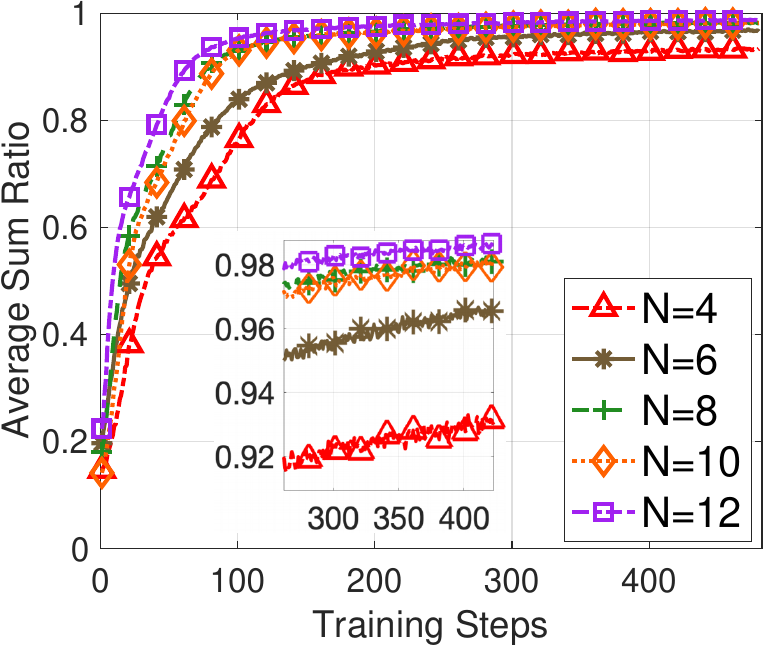}
	\centering
	\caption{Train Set Convergence Performance Comparison with Different Number of Layers}
	\label{fig:mrs_layer}
\end{figure}

% \begin{figure}[htbp]
%   \centering
%   \begin{subfigure}[b]{0.23\textwidth}
%     \includegraphics[width=\textwidth]{test1.pdf}
%     \caption{Layer Number Train Set Performance Comparison}
%     \label{fig:layer}
%   \end{subfigure}
%   \hfill 
%   \begin{subfigure}[b]{0.23\textwidth}
%     \includegraphics[width=\textwidth]{train_last_ex4.eps}
%     \caption{Parameters Bound Train Set Performance comparison}
%     \label{fig:bound}
%   \end{subfigure}
  
%   \caption{Hyper-parameters Performance Comparison}
%   \label{fig:imp_para_test}
% \end{figure}

%\subsection{Network Training Effectiveness}
Firstly, we evaluate network training effectiveness. 

Fig. \ref{fig:mrs_layer} shows the evaluation results on network training effectiveness with different number of DU layers. In the figure, the y-axis is the average sum ratio (ASR), defined as $\frac{1}{Q}\sum_{q=1}^Q \frac{\hat{\text{WSR}}_{q,N}}{{\text{WSR}}_{q,N}^*}$. It provides the average performance ratio of DU $\hat{\text{WSR}}$ and optimal $\text{WSR}^*$. 
The DU performance is closer to optimality then the corresponding ASR approaches 1. From the evalution results we can see that the proposed algorithm can almost reach 98\% ASR when layer number is 12 after just a few training steps. As the number of layers increases, ASR  continues to increase but the marginal improvement becomes negligible after $N=8$. It is because each layer in DU is constructed according to PGD iteration, which follows the iterative gradient descent process in optimization algorithm. As $N$ grows, DU is also progressed to better results due to model convergence. 

% In Table \ref{table:important_parameters_comparison}, DU's performance in test set is provided with different hyper-parameters, which maintains its generalization ability. This explains that although the DU is trained by a small data set, it can also converge rapidly to near-optimal results. The best ASR can reach over 97\% with hyper-parameters: $N=12, lr=0.003, bz=200$.

% Especially, in the parameters bound design test, the comparison for threshold $\epsilon$ is provided which is designed to enforce the parameters to follow gradient updating direction without over-fitting. Otherwise, if $\epsilon$ does not exist, performance drops significantly.

\subsection{Out-of-distribution Test}

%In this subsection, we apply the same hyper-parameters setting for OOD test. 

%Since data-driven neural networks perform much worse in OOD \cite{shlezinger2023model}, while model-driven is designed with explicit expression to adapt to various OOD scenarios. 

We test the generalizability on two OOD scenarios which are: different channel SNR (Scenario 1); different $\text{P}_\text{max}$ (Scenario 2). Note that the same hyper-parameter settings are used for OOD test. In Scenario 1, the channel SNR is changed from the original $40$ dB (for training) to $35$ dB and $45$ dB (for testing). For the second scenario, $\text{P}_\text{max}$ is changed from $33$ dBm (for training) to $32$ dBm and $34$ dBm (for testing), respectively. Since the learnable parameters are restricted by the explicit expression computation in DU networks, the DU can maintain its performance in training set and learn the optimization trajectory underlying optimization rules. As depicted in Table \ref{table:generization_ability}, both two scenarios can reach an ASR around $95\%$ or better. Therefore, DU shows comparable ASR performance to the baseline, which demonstrates its strong resilience to more general communication settings.

\begin{table}[ht]
\centering
\caption{OOD Performance Results}
\begin{tabular}{|c|c|c|c|} 
\hline
{Channel SNR}& Original SNR $40$ dB & $35$ dB &  $45$ dB \\
\hline
ASR (\%) & 98.25 & 94.43  & 95.39 \\ 
\hline
{Maximum Power}& Original $\text{P}_\text{max}$ $33$ dB& $32$ dB &  $34$ dB \\
\hline
ASR (\%) & 98.25  & 96.50  & 97.25 \\ 
\hline

\end{tabular}

\label{table:generization_ability}
\end{table}%
\subsection{Complexity Comparison with FP-based Optimization}

\begin{figure}[ht!]
    \centering
    \includegraphics[width=2.9in]{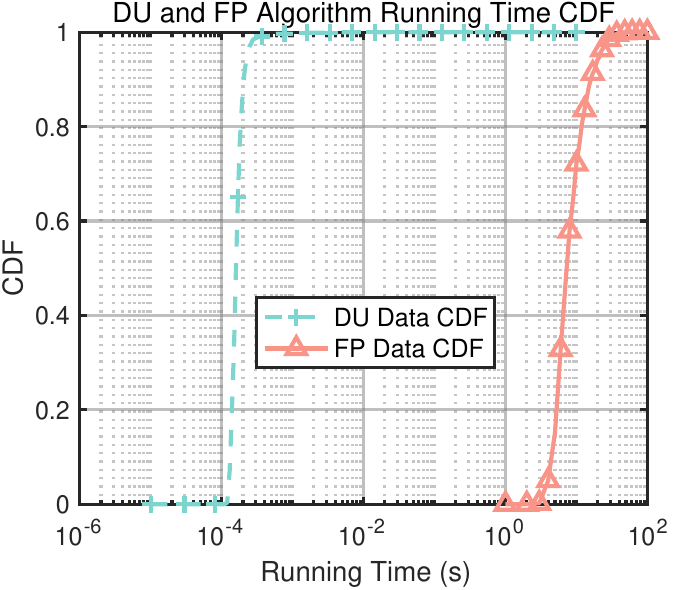}
	\centering
	\caption{Running time CDF of FP and proposed DU}
	\label{CDF_Computation_complexity}
\end{figure}

In this subsection, we analyze and compare the computational complexity of the proposed method with the traditional optimization approach. In specific, we implement the DU trained parameters from Python on MATLAB and setup a DU forward propagation structure with $N=12$ layers, to compare the execution time with FP algorithm given by CVX tool box.  We repeat Monte Carlo experiments of FP and proposed DU for 1,000 times to obtain the execution time distribution cumulative distribution function~\ref{CDF_Computation_complexity}. Our evaluation results show that the proposed DU method is 3,000 times faster than the traditional FP-based optimization on average, where the mean value of DU running time and FP running time are 0.0028 and 10.1569 seconds, respectively. Moreover, since the number of layers in DU is fixed, the computation amount is fixed. In other words, the execution time is bounded within a certain range. Therefore, this method can well fit those communication scenarios that require low latency. 
% \begin{table}[ht]
% \centering
% \caption{Computation Time  Comparison}
% \begin{tabular}{|c|c|c|} 
% \hline
% Approach & FP-based Optimization & Proposed DU\\ 
% \cline{1-3} 
% Total Time & 8.17392& $\mathbf{0.05232}$ \\
% \hline
% Time for Each Step/ Layer & 2.04348 & $\mathbf{0.00436}$ \\
% \hline

% \end{tabular}

% \label{table:complexity}
% \end{table}

\section{Conclusion}\label{section_V}
This paper focused on the downlink  WSR multi-user communication  problem  with RSMA. We aimed to obtain the optimal beamforming vectors for each user. By applying a well-designed penalty function, we proposed an FP and PGD based DU neural network framework, which shows comparable performance with traditional optimization approach but with much lower complexity. This paper provided extensive numerical results and has shown the advantages of proposed method in ASR, OOD, and computation time performance. 
%Besides, proposed DU performs great generalizability in test set with $96.71\%$ ASR while $96.65\%$ ASR in train set. Besides, it still maintains at least $91.55\%$ ASR in OOD test with 156 times faster than the traditional IPM algorithm. Therefore, this framework has great potential to be applied in many scenarios, especially where latency requirement is high.

\bibliographystyle{IEEEtran} 
\bibliography{lib}

\begin{thebibliography}{10}
\providecommand{\url}[1]{#1}
\csname url@samestyle\endcsname
\providecommand{\newblock}{\relax}
\providecommand{\bibinfo}[2]{#2}
\providecommand{\BIBentrySTDinterwordspacing}{\spaceskip=0pt\relax}
\providecommand{\BIBentryALTinterwordstretchfactor}{4}
\providecommand{\BIBentryALTinterwordspacing}{\spaceskip=\fontdimen2\font plus
\BIBentryALTinterwordstretchfactor\fontdimen3\font minus \fontdimen4\font\relax}
\providecommand{\BIBforeignlanguage}[2]{{%
\expandafter\ifx\csname l@#1\endcsname\relax
\typeout{** WARNING: IEEEtran.bst: No hyphenation pattern has been}%
\typeout{** loaded for the language `#1'. Using the pattern for}%
\typeout{** the default language instead.}%
\else
\language=\csname l@#1\endcsname
\fi
#2}}
\providecommand{\BIBdecl}{\relax}
\BIBdecl

\bibitem{clerckx2023primer}
B.~Clerckx, Y.~Mao, E.~A. Jorswieck, J.~Yuan, D.~J. Love, E.~Erkip, and D.~Niyato, ``A primer on rate-splitting multiple access: Tutorial, myths, and frequently asked questions,'' \emph{IEEE Journal on Selected Areas in Communications}, 2023.

\bibitem{clerckx2016rate}
B.~Clerckx, H.~Joudeh, C.~Hao, M.~Dai, and B.~Rassouli, ``Rate splitting for mimo wireless networks: A promising phy-layer strategy for lte evolution,'' \emph{IEEE Communications Magazine}, vol.~54, no.~5, pp. 98--105, 2016.

\bibitem{yang2021optimization}
Z.~Yang, M.~Chen, W.~Saad, and M.~Shikh-Bahaei, ``Optimization of rate allocation and power control for rate splitting multiple access (rsma),'' \emph{IEEE Transactions on Communications}, vol.~69, no.~9, pp. 5988--6002, 2021.

\bibitem{yang2020sum}
Z.~Yang, M.~Chen, W.~Saad, W.~Xu, and M.~Shikh-Bahaei, ``Sum-rate maximization of uplink rate splitting multiple access (rsma) communication,'' \emph{IEEE Transactions on Mobile Computing}, vol.~21, no.~7, pp. 2596--2609, 2020.

\bibitem{xiao2023joint}
M.~Xiao, H.~Cui, Z.~Zhao, X.~Cao, and D.~O. Wu, ``Joint 3d deployment and beamforming for rsma-enabled uav base station with geographic information,'' \emph{IEEE Transactions on Wireless Communications}, 2023.

\bibitem{dizdar2023rsma}
O.~Dizdar, A.~Sattarzadeh, Y.~X. Yap, and S.~Wang, ``Rsma for overloaded mimo networks: Low-complexity design for max-min fairness,'' \emph{IEEE Transactions on Wireless Communications}, 2023.

\bibitem{ma2024exploring}
X.~Ma, Y.~Zhou, H.~Zhang, Q.~Wang, H.~Sun, H.~Wang, and R.~Q. Hu, ``Exploring communication technologies, standards, and challenges in electrified vehicle charging,'' \emph{arXiv preprint arXiv:2403.16830}, 2024.

\bibitem{clerckx2024multiple}
B.~Clerckx, Y.~Mao, Z.~Yang, M.~Chen, A.~Alkhateeb, L.~Liu, M.~Qiu, J.~Yuan, V.~W. Wong, and J.~Montojo, ``Multiple access techniques for intelligent and multi-functional 6g: Tutorial, survey, and outlook,'' \emph{arXiv preprint arXiv:2401.01433}, 2024.

\bibitem{zhang2022deep}
Q.~Zhang, L.~Zhu, S.~Jiang, and X.~Tang, ``Deep unfolding for cooperative rate splitting multiple access in hybrid satellite terrestrial networks,'' \emph{China Communications}, vol.~19, no.~7, pp. 100--109, 2022.

\bibitem{gao2022online}
J.~Gao, C.~Zhong, G.~Y. Li, and Z.~Zhang, ``Online deep neural network for optimization in wireless communications,'' \emph{IEEE Wireless Communications Letters}, vol.~11, no.~5, pp. 933--937, 2022.

\bibitem{zhang2023map2schedule}
L.~Zhang, H.~Sun, J.~Sun, R.~Parasuraman, Y.~Ye, and R.~Q. Hu, ``Map2schedule: An end-to-end link scheduling method for urban v2v communications,'' 2023.

\bibitem{shlezinger2023model}
N.~Shlezinger, J.~Whang, Y.~C. Eldar, and A.~G. Dimakis, ``Model-based deep learning,'' \emph{Proceedings of the IEEE}, 2023.

\bibitem{schynol2023coordinated}
L.~Schynol and M.~Pesavento, ``Coordinated sum-rate maximization in multicell mu-mimo with deep unrolling,'' \emph{IEEE Journal on Selected Areas in Communications}, vol.~41, no.~4, pp. 1120--1134, 2023.

\bibitem{zhang2023joint}
J.~Zhang, C.~Masouros, and L.~Hanzo, ``Joint precoding and csi dimensionality reduction: An efficient deep unfolding approach,'' \emph{IEEE Transactions on Wireless Communications}, 2023.

\bibitem{gallier2019fundamentals}
J.~Gallier and J.~Quaintance, ``Fundamentals of optimization theory with applications to machine learning,'' \emph{University of Pennsylvania Philadelphia, PA}, vol. 19104, 2019.

\bibitem{xia2023deep}
W.~Xia, Y.~Jiang, B.~Zhao, H.~Zhao, and H.~Zhu, ``Deep unfolded fractional programming based beamforming in ris-aided miso systems,'' \emph{IEEE Wireless Communications Letters}, 2023.

\bibitem{shen2018fractional}
K.~Shen and W.~Yu, ``Fractional programming for communication systems—part i: Power control and beamforming,'' \emph{IEEE Transactions on Signal Processing}, vol.~66, no.~10, pp. 2616--2630, 2018.

\bibitem{grant2014cvx}
M.~Grant and S.~Boyd, ``Cvx: Matlab software for disciplined convex programming, version 2.1,'' 2014.

\bibitem{samuel2019learning}
N.~Samuel, T.~Diskin, and A.~Wiesel, ``Learning to detect,'' \emph{IEEE Transactions on Signal Processing}, vol.~67, no.~10, pp. 2554--2564, 2019.

\bibitem{nguyen2020deep}
N.~T. Nguyen and K.~Lee, ``Deep learning-aided tabu search detection for large mimo systems,'' \emph{IEEE Transactions on Wireless Communications}, vol.~19, no.~6, pp. 4262--4275, 2020.

\end{thebibliography}

\end{document}